\documentclass[12pt,prd,superscriptaddress,preprintnumbers,amsmath,amssymb,nofootinbib]{revtex4-2}
\usepackage{epsfig}  
\usepackage{graphicx}
\usepackage{hyperref}
\usepackage{color}
\usepackage{float}
\usepackage{amsfonts}
\usepackage{amsmath}
\usepackage{slashed}
\usepackage{soul}
\usepackage{braket}
\usepackage{multirow}
\usepackage[utf8]{inputenc}
\usepackage[table]{xcolor}
\usepackage{appendix}

\newcommand{\sstk}{\sin^2\theta_{K^*}}
\newcommand{\sstl}{\sin^2\theta_{\ell}}
\newcommand{\stk}{\sin\,2\theta_{K^*}}
\newcommand{\stl}{\sin\,2\theta_{\ell}}
\newcommand{\cstk}{\cos^2\theta_{K^*}}
\newcommand{\ctl}{\cos\,2\theta_{\ell}}

\large

\begin{abstract}
\section*{Abstract}
We analyze the new physics effects in semileptonic $\bar{B_s} \to K^{*+}(\to K\pi) \ell^- \bar{\nu}_\ell$ decay induced by the $b \to u \ell \nu_{\ell}$ quark level transition. We consider the vector, scalar and tensor new physics Lorentz structures in addition to the SM in effective field theory approach. Constraints on new physics parameters are obtained from experimental observations of both semileptonic and leptonic decays of $B$ mesons, which are governed by the underlyng $b \to u \ell \nu_\ell$ transition. We explore the new physics effects in differential branching fraction, lepton forward-backward asymmetry, fraction of longitudinal polarization of $K^*$ meson and normalized angular observables in $\bar{B_s} \to K^{*+}(\to K\pi) \ell^- \bar{\nu}_\ell$ decay.
\end{abstract}

\begin{document} 
\preprint{}

\title{New physics effects in semileptonic $\bar{B_s} \to K^{*+}(\to K\pi) \ell^- \bar{\nu}_\ell$ decay }

\author{Shabana Khan}
\email{shabana26k@gmail.com}
\affiliation{Department of Physics, University of Rajasthan, Jaipur 302004, India}

\author{Dinesh Kumar}
\email{dineshsuman09@gmail.com(corresponding author)}
\affiliation{Department of Physics, University of Rajasthan, Jaipur 302004, India}

\maketitle

\section{Introduction} 
The Standard Model (SM) of particle physics provides a detailed description about the fundamental interactions of nature. However, it is widely acknowledged that the SM is not the ultimate framework for describing the fundamental nature of the universe, leaving room for the new phenomena beyond its framework. There are several phenomena, such as baryon asymmetry, neutrino oscillations, and the existence of dark matter that cannot be explained within the Standard Model. This indicates the new physics (NP) beyond the SM. No direct evidence has yet emerged from high-energy particle colliders to confirm the existence of any new particle beyond those predicted by the Standard Model. Another way to search for new physics is to look for indirect evidence through precision measurements in low-energy processes.  

The semileptonic decays offer a promising avenue for exploring the new physics beyond the SM. The semileptonic $B$ decays have attracted a lot of attention of high energy community over the recent years. The investigation of B meson decays has played a crucial role for testing SM predictions and exploring potential NP beyond it. These decays are instrumental for providing the nature of NP interaction beyond the SM. The semileptonic decays of $B$ mesons with flavor changing charged current (FCCC), where a quark changes the flavor through the exchange of $W$-boson,  have provided the hints for NP. Recent measurements from Belle, LHCb and BaBar have provided the intriguing hints of the deviation from the SM predictions.

The few unexpected deviations are observed in the semileptonic B decay induced by charged currents $b \to c \ell \nu_{\ell}$. The lepton flavor universality observable is defined as the ratio $R_{D^{(*)}} = \frac{\mathcal{B}(B \to D^{(*)}\tau \nu)}{\mathcal{B}(B \to D^{(*)}\ell \nu)}$, with $\ell = e, \mu$.  After a new measurement from LHCb \cite{LHCb:2024jll}, the new world average of this ratio is in tension with SM at the level of $3.31\, \sigma$ \cite{hflav}. The world average is the average of measurements from different experiments BABAR\cite{BaBar:2012obs,BaBar:2013mob}, Belle\cite{Huschle:2015rga,Hirose:2016wfn,Belle:2017ilt,Belle:2019rba}, Belle II\cite{Belle-II:2024ami} and LHCb \cite{LHCb:2023zxo,LHCb:2023uiv,LHCb:2024jll}. The deviation of the world average of measured values from SM predictions provides a hint for LFU violation in tau and light leptons, which indicates the presence of NP. Belle collaboration has provided the measurement of $\tau$-longitudinal polarization asymmetry $(P_{\tau}^{D^*})$ and $D^*$ polarization fraction observables $(F_L^{D^*})$ in $B \to D^* \tau \nu$ \cite{Hirose:2016wfn, Belle:2017ilt,Belle:2019ewo}. The measured values of these observables are $P_{\tau}^{D^*} = -0.38 \pm 0.51 (stat) ^{+0.21}_{-0.16} (sys)$ and $F_L^{D^*} =  0.60 \pm 0.08 (stat) \pm 0.04 (syst)$. These measurements agree with SM predictions within $\sim 1.5 \sigma$ \cite{Alok:2016qyh, Iguro:2020cpg}. The LFU ratio $R_{J/\Psi}$ in charmed meson decay $B_c \to J/\Psi \ell \nu$ was analyzed by LHCb collaboration, and the experimentally measured value of this observable is $R_{J/\Psi}^{\tau/\mu} = 0.71 \pm 0.17 \pm 0.18$ \cite{LHCb:2017vlu} which is consistent with SM within $1.8 \sigma$ \cite{Harrison:2020nrv}.

These discrepancies in the measurements from SM can be explained by considering the new physics effects in extensions of SM with new interactions and can be found in a very recent global fit work \cite{Iguro:2024hyk} and references therein. However, $b \to u \ell \nu$ decays are suppressed due to the CKM matrix element $V_{ub}$ compared to the $b \to c \ell \nu$. The inclusive and exclusive $B$-hadron decays have long produced conflicting results for the Cabibbo-Kobayashi-Maskawa (CKM) matrix element $|V_{ub}|$\cite{HFLAV:2022esi,Ricciardi:2019zph,Ricciardi:2021shl,Capdevila:2021vkf}, which may point to the possibility of new physics. The SM parameter $|V_{ub}|$ is determined by semileptonic $b \to u$ transitions. Semileptonic $B$ decays continue to provide a powerful testing ground for the SM and possible signals of NP, especially in the context of LFU. While measurements of $b \to c \tau \nu$ transitions have shown persistent tensions with SM predictions, current data on $b \to u\ell\nu$ decays involving light leptons ($\ell = e, \mu$) exhibit no significant deviations. The authors in the ref. \cite{Greljo:2023bab} discuss that even without significant tensions between measurements and SM predictions, $b \to u \ell \nu_{\ell}$ processes provide valuable insights and constraints on new physics models. The central values of $V_{ub}$ extracted from the $B \to \rho \ell \nu$ and $B \to \omega \ell \nu$ is found to be smaller than the value extracted from the inclusive decay or $B \to \pi \ell \nu $ decay\cite{Bernlochner:2021rel}. The precise exclusive analysis in light lepton channels directly probes the size and nature of any LFU/LFUV NP contribution.  The new physics in $b \to u \ell \nu$ decays have been investigated in weak effective field theory in few more studies \cite{Duan:2024ayo,Leljak:2023gna,Rajeev:2018txm,Dutta:2016eml}. 

The transitions involving $b \to u \ell \nu$ provide an avenue to testing the structure of SM and potential NP. The semileptonic transitions $b \to u$ are important due to their role in determining the CKM matrix element, which governs the coupling strength between $b$ quark and $u$ quark. The precise measurement of CKM matrix element $V_{ub}$ is important for testing the unitarity of the CKM matrix. The discrepancies between different inclusive and exclusive measurements of $V_{ub}$ have raised question whether unknown new physics interactions can affect this transition.

The study of rare decays in the realm of flavor physics offers a unique window into potential new physics beyond the Standard Model (SM). Among these decays, the  transition $\bar{B_s} \to K^{*+}(\to K\pi) \ell^- \bar{\nu}_{\ell}$ stands out as a particularly promising avenue for probing physics beyond the SM. In this work, we analyzed the allowed new physics constrained by the available $b \to u \ell\nu_{\ell}$ data. We considered the different NP Lorentz structures in weak effective field theory. This decay is very useful for extraction of CKM element $V_{ub}$. We have utilized the available experimental measurement for leptonic decay $B \to \mu \nu_{\mu}$ \cite{Belle:2019iji} from Belle collaboration. In addition to this leptonic decay measurement, we have also used the measurements from BaBar and Belle collaborations for semileptonic decays $B \to (\pi, \rho, \omega) \ell \nu_{\ell}$  and globally averaged $q^2$ binned spectrum for differential decay width of these decays are provided by HFLAV collaboration \cite{HFLAV:2022esi}. These experimental measurements can provide the information about the NP sensitivity with the light leptons.

We aim to provide a comprehensive analysis of $\bar{B_s} \to K^{*+}(\to K\pi) \ell^- \bar{\nu}_\ell$ decay process, focusing particularly on its sensitivity to new physics effects. We present theoretical expressions of the differential branching fraction, leptonic forward-backward asymmetry and longitudinal polarization fraction of $K^*$ meson in this decay within the SM framework and with the NP interactions. We also defined the normalized angular observables which are independent to the CKM element $V_{ub}$. The future experimental measurements for this decay will be very helpful to provide crucial hints about the existence and nature of new physics phenomena. This decay has been investigated in ref. \cite{Feldmann:2015xsa} discussing the usefulness of the angular observables  and the synergy between the observables $\bar{B} \to \bar{K}^* \ell^+ \ell^-$ and $\bar{B_s} \to K^{*+}(\to K\pi) \ell^- \bar{\nu}_{\ell}$. We have studied in detail with the possible 1D and 2D scenarios with different NP Lorentz structures. The LHCb experiment, with its high-statistics environment and excellent capabilities for flavor tagging and vertex reconstruction, is expected to enable the first precision measurements of the differential decay rate and angular observables in the decay $\bar{B_s} \to K^{*+}(\to K\pi) \ell^- \bar{\nu}_\ell$ during Run 3. Such measurements would be crucial for constraining new physics via global fits to the Wilson coefficients of the relevant effective operators~\cite{LHCb:2018roe}. Although Belle~II faces limited $B_s$ production at the $\Upsilon(5S)$, its clean experimental environment and nearly hermetic detector offer complementary sensitivity especially to branching fractions and lepton flavor universality observables such as $R_{K^*}^{B_s}$ provided sufficient data are collected at the $\Upsilon(5S)$ resonance~\cite{Belle-II:2018jsg}.\\
In this work, we use a general parametrization of the effective Hamiltonian to study the decay \( B_s \to K^{*} \ell \bar{\nu} \). We note that in the Standard Model Effective Field Theory (SMEFT), the coefficients in the Hamiltonian follow specific rules due to the structure of the electroweak interaction. While we do not apply these SMEFT rules in our current work, including them in future studies could improve the accuracy of our results and provide a more detailed analysis. Our work is organized as follows : In section \ref{framework}, we have described the theoretical framework for $b \to u \ell \nu_{\ell}$ processes in effective field theory framework with new physics. In section \ref{par_space}, we have studied the allowed NP parameter space using the available experimental measurements in this sector. In section \ref{distribution}, the angular distribution of $\bar{B_s} \to K^{*+}(\to K\pi) \ell^- \bar{\nu}_l$ is defined within the SM and having NP effects. We provide the predictions for the $q^2$ spectrum of differential branching fraction, leptonic forward-backward asymmetry, longitudinal polarization fraction of $K^*$ meson and normalized angular observables in this decay in section \ref{predictions}. The conclusion of our analysis is summarized in section \ref{summary}. The details of form factors and helicity amplitudes for $\bar{B_s} \to K^{*+}(\to K\pi) \ell^- \bar{\nu}_l$ are given in the appendices.

\section{Theoretical framework}
\label{framework}
We consider the general low-energy effective Hamiltonian for the $b \to u \ell\nu_\ell$ transitions with possible different new physics Lorentz structure contributions given by
\begin{equation}
\mathcal{H}_{eff}^{b \to u} = -\frac{4G_F}{\sqrt{2}}V_{ub} \left[\,(1 + C_{V_L})\,\mathcal{O}_{V_L} + C_{V_R}\, \mathcal{O}_{V_R} + C_{S_L}\, \mathcal{O}_{S_L}+ C_{S_R}\, \mathcal{O}_{S_R}+ C_{T}\, \mathcal{O}_{T} \right] + h.c.
\end{equation}
where $G_F$ and $V_{ub}$ are the Fermi coupling constant and CKM matrix element, respectively. The operators $O_i$'s are four fermion operators and $C_i$'s are the short-distance Wilson coefficients. The dimension-6 operators $O_i$'s are given as
\begin{eqnarray}
\nonumber \mathcal{O}_{V_L} & = & (\bar{u}\gamma^{\mu}P_L b)(\bar{\ell}\gamma_{\mu}P_L \nu_{\ell}), \quad \mathcal{O}_{V_R} = (\bar{u}\gamma^{\mu}P_R b)(\bar{\ell}\gamma_{\mu}P_L \nu_{\ell})\\ \nonumber
\mathcal{O}_{S_L} &=& (\bar{u}P_L b)(\bar{\ell}P_L \nu_{\ell}), \quad\quad\quad \mathcal{O}_{S_R} = (\bar{u}P_R b)(\bar{\ell}P_L \nu_{\ell})\\  
\mathcal{O}_{T} &=& (\bar{u}\sigma^{\mu\nu}b)(\bar{\ell}\,\sigma_{\mu\nu}P_L \nu_{\ell}), 
\end{eqnarray}
We exclusively consider neutrinos as left-handed particles. The vector $C_{V_{L,R}}$, scalar $C_{S_{L,R}}$, and tensor $C_T$ Wilson coefficients encode the contribution from NP Lorentz structures. In our analysis, we consider the NP contribution to be real only to focus on observables that are currently measured with the best precision, which are primarily CP-conserving. We assume the lepton flavor universal NP couplings for light leptons, i.e. electron and muon. Current experimental data on semileptonic decays involving light leptons (\( \ell = e, \mu \)) show no statistically significant deviation from lepton flavor universality in the \( b \to u \ell \nu \) sector. In particular, the Belle measurement of \( B \to D \ell \nu \)~\cite{Belle:2015pkj} demonstrates comparable experimental sensitivity in the electron and muon channels, supporting the assumption of LFU. While a global EFT fit separating the two light lepton flavors in \( b \to u \ell \nu \) is still limited by statistics, existing analyses\cite{Greljo:2023bab,Leljak:2023gna} treat the NP Wilson coefficients as lepton-flavor universal for \( \ell = e, \mu \), a convention we adopt in this work.  We consider the NP couplings as $C_i^{\ell} = (C_i^e + C_i^{\mu})/2$ as the most of the available data are the combination of electron and muon channel.

We consider the following measurements to constrain the NP couplings:
\begin{itemize}
\item For the decay mode $\bar{B}^0 \to \pi^+\ \ell^- \bar{\nu_{\ell}}$, we have utilized the globally averaged $q^2-$ binned branching ratio spectrum published by the HFLAV collaboration \cite{HFLAV:2022esi}. This average incorporates data from the BaBar \cite{BaBar:2010efp, BaBar:2012thb} and Belle \cite{Belle:2010hep, Belle:2013hlo} experiments and systematic correlations among their individual results are taken into account. 
\item We use the world average of the differential branching fractions in different $q^2$ bins for the decays $B \to \rho \ell \nu_{\ell}$ and $B \to \omega \ell \nu_{\ell}$ published by the HFLAV collaboration\cite{HFLAV:2022esi}. The average for $B \to \rho \ell\nu_{\ell}$ is obtained from BaBar\cite{BaBar:2010efp} and Belle\cite{Belle:2013hlo} measurements. The experimental measurements are also available from Belle\cite{Belle:2013hlo} and BaBar\cite{BaBar:2012dvs} for $B \to \omega \ell \nu_{\ell}$ which were used to obtain the average spectrum for this decay.
\item The measurement of the branching ratio of the leptonic decay $\mathcal{B}(B^- \to \mu^-\,\bar{\nu}_{\mu})$ \cite{Belle:2019iji} from the Belle collaboration is also used to constrain the NP parameters.
\end{itemize}

The value of $V_{ub}$ is determined from measured partial branching fractions, utilizing a variety of theoretical predictions for the decay rate. In principle, the inclusive decay rate for $B \to X_u \ell \nu$ can be computed using the heavy quark expansion (HQE), by expanding in powers of $1/m_b$\cite{Belle:2021eni}. The analysis of inclusive $B \to X_u \ell \nu$ decays is hindered by significant uncertainties, largely stemming from the need to subtract the substantial $B \to X_c \ell \nu$ background. Moreover, the background suppression strategies involve stringent phase-space cuts, which in turn limit the reliability of the theoretical description based on the heavy quark expansion \cite{Lange:2005yw, Gambino:2007rp, Andersen:2005mj}. Given these challenges, we do not consider inclusive decays in this work and focus mainly on exclusive $b \to u \ell \nu$ transitions, where both theoretical control and experimental precision are more favorable.

The leptonic decay $B \to \mu\nu_{\mu}$ is induced by $b \to u\, \mu\,\nu_{\mu}$ transition and the branching ratio of this decay can be expressed in terms of NP WCs as:
\begin{equation}
\frac{\mathcal{B}(B \to \mu\nu_{\mu})^{NP}}{\mathcal{B}(B \to \mu\nu_{\mu})^{SM}} = \Big|1 - C_A^{\mu} + \frac{m_B^2}{m_{\mu}(m_b + m_u)}C_P^{\mu}\Big|^2
\end{equation}
where $C_A^{\ell} = C_{V_R}^{\ell} - C_{V_L}^{\ell}$ and $C_P^{\ell} = C_{S_R}^{\ell} - C_{S_L}^{\ell}$ are axial vector and pseudoscalar NP couplings, respectively. The leptonic decay mode is not sensitive to the scalar and tensor NP couplings.

The differential decay rate of semileptonic decay of B meson to a pseudoscalar meson is sensitive to the NP Lorentz structures and the expression in terms of these NP WCs can be written as \cite{Tanaka:2012nw, Sakaki:2013bfa}

\begin{eqnarray}
\frac{d\Gamma(B \to P\, \ell\,\nu_{\ell})}{dq^2} & = & \frac{G_F^2|V_{ub}|^2 q^2}{192 \pi^3 m_B^3}\sqrt{\lambda_P(q^2)}\Big(1-\frac{m_l^2}{q^2}\Big)^2 \times \Big\{ \nonumber \\
&&\Big| 1 + C_{V_L}^{\ell} + C_{V_R}^{\ell}\Big|^2 \Big[ \Big(1+\frac{m_{\ell^2}}{2q^2}\Big)H_{V,0}^{s^2} + \frac{3}{2}\frac{m_{\ell^2}}{q^2}H_{V,t}^{s^2}\Big] \nonumber \\
&& + \frac{3}{2}|C_{S_L}^{\ell} + C_{S_R}^{\ell}|^2\, H_S^{s^2} + 8|C_T^{\ell}|^2 (1+ \frac{2m_{\ell^2}}{q^2})H_T^{s^2} \nonumber \\
&& +  3\, Re[(1 + C_{V_L}^{\ell} + C_{V_R}^{\ell})\,(C_{S_L}^{\ell*} + C_{S_R}^{\ell*})]\frac{m_{\ell}}{\sqrt{q^2}}H_S^s\, H_{V,t}^s \nonumber\\
&& - 12 \, Re[(1 + C_{V_L}^{\ell} + C_{V_R}^{\ell})\,C_T^{\ell*}]\frac{m_\ell}{\sqrt{q^2}}H_T^s\, H_{V,0}^s \Big\}
\end{eqnarray}

where $\lambda_P = ((m_B -m_P)^2 - q^2)((m_B + m_P)^2 -q^2)$ and $H_V^s, H_S^s$ and $H_T^s$ are the helicity amplitudes. The expressions of these helicity amplitudes are provided in ref. \cite{Sakaki:2013bfa}. The semileptonic decay $B \to \pi\, \ell\,\nu_{\ell}$ is sensitive to vector $C_V^{\ell}$, scalar $C_S^{\ell}$ and tensor $C_T^{\ell}$ NP couplings. The hadronic matrix elements for this decay can be written in terms of the form factors $f_0(q^2), f_+(q^2), f_T(q^2)$ and the explicit expressions for these form factors are given in ref. \cite{Sakaki:2013bfa}. We have used the form factors calculated in ref. \cite{Leljak:2021vte} where combined fits to lattice QCD data and light cone sum rules are performed.

The differential decay rate of semileptonic decay of B meson to vector meson can be written in terms of NP WCs as \cite{Tanaka:2012nw, Sakaki:2013bfa}
\begin{eqnarray}
\frac{d\Gamma(B \to V\, \ell\,\nu_{\ell})}{dq^2} & = &\frac{G_F^2|V_{ub}|^2 q^2}{192 \pi^3 m_B^3}\sqrt{\lambda_V(q^2)}\Big(1-\frac{m_l^2}{q^2}\Big)^2 \times \Big\{ \nonumber \\
&& \Big(| 1 + C_{V_L}^\ell |^2 + |C_{V_R}^\ell|^2\Big) \Big[ \Big(1+\frac{m_\ell^2}{2q^2}\Big)(H_{V,+}^2 + H_{V,-}^2 + H_{V,0}^2) + \frac{3}{2}\frac{m_\ell^2}{q^2}H_{V,t}^2\Big] \nonumber \\
&& -2\,Re\Big[(1 + C_{V_L}^\ell)C_{V_R}^{\ell*} \Big]\Big[\Big(1 + \frac{m_\ell^2}{2q^2}\Big)(H_{V,0}^2 + 2H_{V,+}H_{V,-}) + \frac{3}{2}\frac{m_\ell^2}{q^2} H_{V,t}^2 \Big] \nonumber \\
&& + \frac{3}{2}|C_{S_R}^\ell - C_{S_L}^\ell|^2\, H_S^2 + 8|C_T^\ell|^2 (1+ \frac{2m_\ell^2}{q^2})(H_{T,+}^2 + H_{T,-}^2 + H_{T,0}^2 ) \nonumber \\
&& +  3\, Re[(1 - C_{V_R}^\ell + C_{V_L}^\ell)\,(C_{S_R}^{\ell*} - C_{S_L}^{\ell*})]\frac{m_\ell}{\sqrt{q^2}}H_S\, H_{V,t} \nonumber\\
&& - 12 \, Re[(1 + C_{V_L}^\ell)\,C_T^{\ell*}]\frac{m_\ell}{\sqrt{q^2}}(H_{T,0}\, H_{V,0} + H_{T,+}\, H_{V,+} - H_{T,-}\, H_{V,-} ) \nonumber \\
&& - 12 \, Re[ C_{V_R}^\ell\,C_T^{\ell*}]\frac{m_\ell}{\sqrt{q^2}}(H_{T,0}\, H_{V,0} + H_{T,+}\, H_{V,+} - H_{T,-}\, H_{V,-} )\Big\}
\end{eqnarray}

The above expression can be used for the semileptonic decays $B \to \rho \ell \nu_{\ell}$ and $B \to \omega \ell \nu_{\ell}$ and the hadronic matrix elements for the $B \to \rho$ and $B \to \omega$ transitions are expressed in terms of the form factors $V(q^2), A_{0,1,2}(q^2)$ and $T_{0,1,2}(q^2)$. The explicit expressions are defined for these hadronic amplitudes in ref. \cite{Sakaki:2013bfa}. 
The form factors for the $B \to \rho$ and $B \to  \omega$ transitions have been computed with the help of light cone sum rules \cite{Bharucha:2015bzk}.

We use the chi-square fit analysis which offers a systematic approach to evaluate the agreement between measured value and expected value. For this purpose, we use the MINUIT library \cite{James:1975dr} which provides a robust and efficient algorithm for minimizing the function with $\chi^2$. Minuit offers functionalities to determine the optimal parameter values and their associated uncertainties by minimizing the $\chi^2$ function. We define the $\chi^2$ for our analysis as
\begin{equation}
\chi^2(C_i)=\sum_{m,n}\left(\mathcal{O}^{th}(C_i)-\mathcal{O}^{exp}\right)_{m} \mathcal{C}^{-1}_{mn}\left(\mathcal{O}^{th}(C_i)-\mathcal{O}^{exp}\right)_{n}
\end{equation}
where $\mathcal{C}^{-1}_{mn}$ is the covariance matrix which includes both experimental and theoretical uncertainties with taking care of the correlations. $\mathcal{O}^{exp}$ and $\mathcal{O}^{th}$ are the experimental measurement and theoretical prediction of the observable, respectively. The theoretical predictions for the observables used in the fit are computed using {\tt flavio}~\cite{flavio}.

\section{Allowed NP parameter space}
\label{par_space}
We consider the 1D and 2D scenarios:
\begin{itemize}
\item {\bf 1D scenarios} : $S1 : C_{V_L},\,\, S2 : C_{V_R},\,\, S3 : C_{S_L},\,\, S4 : C_{S_R},\,\, S5 : C_{T},\,\,S6 : (C_{V_L} = -C_{V_R})$
\item {\bf 2D scenarios} : $S7 : (C_{V_L},C_{V_R}),\,\, S8 : (C_{V_L},C_{S_L}),\,\, S9 : (C_{V_L},C_{S_R}),\,\, S10  : (C_{V_L},C_{T}),\,\, S11 : (C_{V_R},C_{S_L}),\,\, S12 : (C_{V_R},C_{S_R}),\,\, S13 : (C_{V_R},C_{T}),\,\, S14 : (C_{S_L},C_{S_R}),\,\, S15 : (C_{S_L},C_{T}),\,\, S16 : (C_{S_R},C_{T}),\,\, S17 : (C_A, C_P) = (C_{V_L} = -C_{V_R},C_{S_L}=-C_{S_R})$
\end{itemize}

The Standard Model yields a total $\chi^2_{SM} = 24.34$ for 31 observables, corresponding to a reduced chi-square of $\chi^2 /dof = 0.79$.  This reflects a very good agreement between the SM predictions and the experimental data, with no statistically significant tension at present. Nevertheless, we investigate possible new physics scenarios to explore whether certain operator structures can lead to improved fits or reveal patterns that may become significant with future, more precise measurements. Firstly, we allow the NP Lorentz structure one at a time and obtain the best fit values for the NP Wilson coefficients by minimizing the $\chi^2$ function with the available data in $b \to u \ell\,\nu_{\ell}$ sector. The best fit for the 1D NP scenarios are collected in Table \ref{best-fit_1d}.

\begin{table}[htb]
\centering
\begin{tabular}{ |c|c|c|c| }
\hline
\textbf{NP scenario} & \,\,\,\,\,\,\,\,\,\textbf{Best fit}\,\,\,\,\,\,\,\,\, &\,\,\,\,\,\,\,\,\, \textbf{$\chi^2_{min}$} \,\,\,\,\,\,\,&\,\,\,\,\,\,\,\,\, \textbf{$\Delta \chi^2 = \chi^2_{SM} - \chi^2_{min}$}\,\,\,\,\,\,\,\,\,\\
\hline
$S1 : C_{V_L}$ & -0.032(47) &23.87 &0.47 \\
\hline
$S2 : C_{V_R}$  &  0.069(47) &22.31 &2.03 \\
\hline
$S3 : C_{S_L}$  & -0.003(4) &23.86 & 0.48\\
\hline
$S4 : C_{S_R}$ & 0.003(4) &23.86 & 0.48\\
\hline
$S5 : C_{T}$ & 0.005(49) &24.33 & 0.01\\
\hline
$S6 : (C_{V_L} = -C_{V_R})$ &-0.093(54) &20.61 & 3.73\\
\hline

\end{tabular}
\caption{The best fit values of 1D NP scenarios in WET from $b \to u$ data.}
\label{best-fit_1d}
\end{table}

It can be seen from the table~\ref{best-fit_1d} that the NP scenario S6\,($C_{V_L} = -C_{V_R}$) is most preferred by the data, with a $\Delta \chi^2$ of 3.73. This corresponds to a modest ($\sim 2\sigma$) improvement over the Standard Model, suggesting potential hints of new physics, though not yet statistically conclusive. The scenario S2 ($C_{V_R}$) provides the second-best fit with $\Delta \chi^2 = 2.03$, indicating a weaker preference. Other scenarios, including scalar and tensor contributions, show negligible improvement relative to the SM. In particular, the tensor scenario S5 ($C_T$) yields an almost vanishing $\Delta \chi^2 = 0.01$, and is thus disfavored by the data.

We also perform a two-dimensional fit by allowing two NP Wilson coefficients to vary simultaneously. This approach accounts for possible correlations between different operator contributions and provides a more realistic picture of the allowed parameter space. The best-fit values of the NP WCs obtained from these combined scenarios are summarized in Table~\ref{best-fit_2d}, along with the corresponding minimum $\chi^2$ values and $\Delta \chi^2$ improvements over the Standard Model. The correlation between the fitted Wilson coefficients is also shown to illustrate how the parameters influence each other in the fit. These results help to identify whether specific operator combinations can accommodate the data more effectively than single-parameter scenarios.

\begin{table}[htb]
\centering
\begin{tabular}{ |c|c|c|c|c| }
\hline
\,\,\,\,\,\,\,\textbf{NP scenario}\,\,\,\,\,\,\, &\,\,\,\,\,\,\,\,\, \textbf{Best fit} \,\,\,\,\,\,\,\,\,& \textbf{Correlation} &\,\textbf{$\chi^2_{min}$} \,& \,\,\,\,\,\textbf{$\Delta \chi^2 = \chi^2_{SM} - \chi^2_{min}$}\,\,\,\,\,\\
\hline
\multirow{4}{*}{$S7:(C_{V_L}, C_{V_R})$} & $S7a:(-0.079(56),0.115(62))$&\multirow{4}{*}{-0.6} & \multirow{4}{*}{20.22}&\multirow{4}{*}{4.12} \\\cline{2-2}
                                 &$S7b:(-0.892(60),0.928(56))$ &&&\\\cline{2-2}
                                 &$S7c:(-1.122(63),-0.928(57))$ &&& \\\cline{2-2}
                                 &$S7d:(-1.934(58),-0.115(62))$ &&&
                                 \\\cline{2-2}
                                 \hline
$S8:(C_{V_L}, C_{S_L})$&(-0.038(49), -0.003(4))&0.2 & 23.22& 1.12 \\
\hline
$S9:(C_{V_L}, C_{S_R})$ &(-0.038(48),  0.004(4))&-0.2 &23.21 & 1.13\\
\hline
$S10:(C_{V_L}, C_{T})$&(-0.032(47),  0.006(57))&-0.1&23.86&0.48\\
\hline
$S11:(C_{V_R}, C_{S_L})$ & (0.075(48),  -0.004(4)) &-0.2&21.44&2.9 \\
\hline
$S12:(C_{V_R}, C_{S_R})$&(0.075(48),  0.004(4)) &0.2&21.46&2.88\\
\hline
$S13:(C_{V_R}, C_{T})$&(0.068(48),  0.0007(50))&-0.1&22.31&2.03 \\
\hline
$S14:(C_{S_L}, C_{S_R})$ &(0.008(121), 0.011(120))&1.0&23.85& 0.45\\
\hline
$S15:(C_{S_L}, C_{T})$ &(-0.003(4),  0.005(49))  &0.0&23.85&0.45\\
\hline
$S16:(C_{S_R}, C_{T})$ &(0.003(4), 0.005(49))  &0.0&23.85&0.45\\
\hline
$S17:(C_A, C_P)$&(-0.116(59), 0.015(2))&0.4&18.84&5.5\\
                                 \hline
\end{tabular}
\caption{The best fit values of 2D NP scenarios in WET from $b \to u$ data.}
\label{best-fit_2d}
\end{table}

Based on the \(\Delta \chi^2\) values presented in Table~\ref{best-fit_2d}, we assess the relative preference for various NP scenarios compared to the SM. The most favored scenario is \(S17\), which corresponds to the specific combination \((C_A, C_P) = (C_{V_L} = -C_{V_R},\; C_{S_L} = -C_{S_R})\). This scenario achieves the lowest \(\chi^2_{\min}\) and the highest improvement over the SM with \(\Delta \chi^2 = 5.5\). This corresponds to a statistically significant preference (above \(2\sigma\)), suggesting a potentially important role for these operators in explaining the \(b \to u \ell \nu\) data.

The next most preferred scenario is \(S7\), which includes simultaneous contributions from \((C_{V_L}, C_{V_R})\). This scenario yields \(\Delta \chi^2 = 4.12\), with multiple best-fit solutions (S7a–S7d) and a strong negative correlation between the two coefficients, indicating a nontrivial interplay and possible degeneracy structure in the parameter space. Scenarios \(S11\) and \(S12\), involving vector-scalar combinations \((C_{V_R}, C_{S_L})\) and \((C_{V_R}, C_{S_R})\) respectively, also show moderate improvements with \(\Delta \chi^2 \approx 2.9\), suggesting that scalar contributions can be relevant when paired with right-handed vector currents.

A milder preference is observed for scenarios like \(S13\), \((C_{V_R}, C_T)\), which has a \(\Delta \chi^2 = 2.03\), comparable to the one-dimensional \(C_{V_R}\) fit, implying the tensor contribution does not significantly enhance the fit. Similarly, combinations involving scalar operators with \(C_{V_L}\), such as \(S8\) and \(S9\), provide only marginal improvements (\(\Delta \chi^2 \approx 1.1\)). In contrast, scenarios like \(S10\), \(S14\), \(S15\), and \(S16\), which involve scalar-scalar or scalar-tensor combinations, show negligible improvements (\(\Delta \chi^2 < 0.5\)), and are therefore not supported by the current data.

\begin{figure}[htb!]
\includegraphics[height=0.24\textwidth,width=0.3\textwidth]{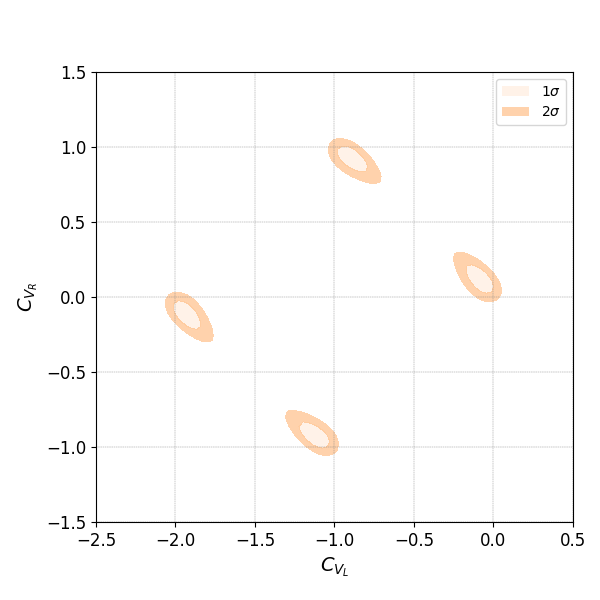}
\includegraphics[height=0.24\textwidth,width=0.3\textwidth]{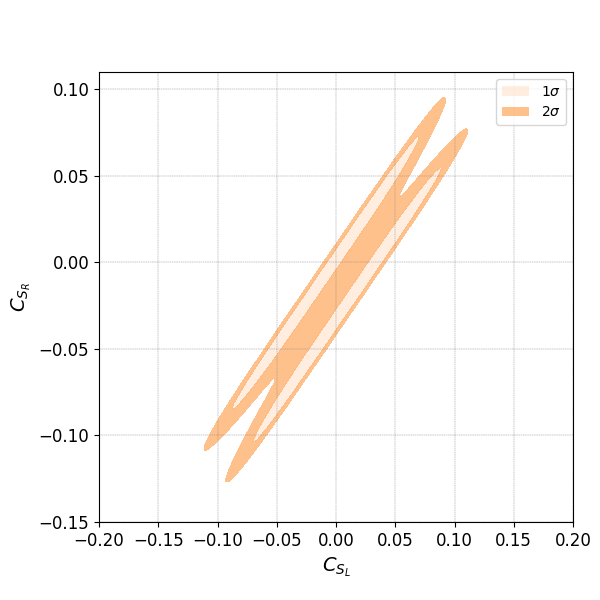}
\includegraphics[height=0.24\textwidth,width=0.3\textwidth]{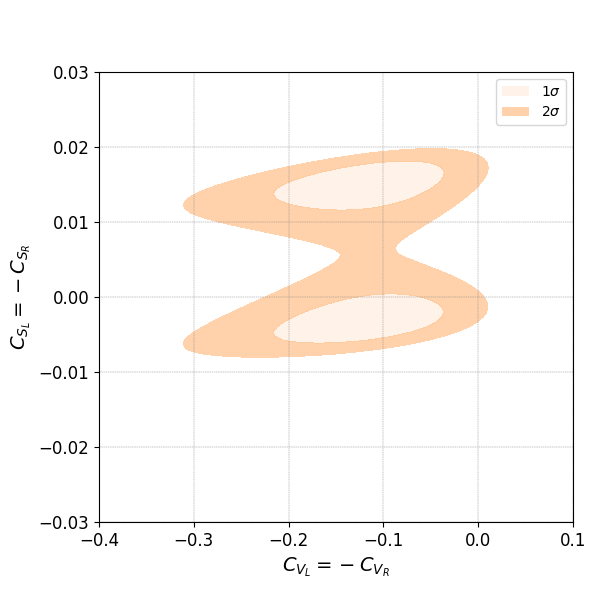}
\caption{Contours for the allowed parameter space of $68\%$ (light orange area) and $95\%$ (orange area) for new Physics WCs obtained from our fit using the available measurements.}
\label{2d_space}
\end{figure}

We plot the $\chi^2$ contours in the plane of two dimensional WCs with $68\%$ and $95\%$ confidence intervals, which are related to the $\Delta \chi^2 = 2.30, 5.99$, respectively, for the two degrees of freedom. The allowed parameter space for scenarios $S7, S14 \, \& \, S17$ are shown in figure \ref{2d_space}.

\section{Angular Analysis in $\bar{B_s} \to K^{*+}(\to K\pi) \ell^- \bar{\nu}_\ell$}
\label{distribution}
The fully differential decay rate for the semileptonic decay $\bar{B_s} \to K^{*+}(\to K\pi) \ell^- \bar{\nu}_\ell$ can be written similar to the decay rate of FCNC decay $B \to K^*(\to K\pi)\ell^+\ell^-$ \cite{Jager:2012uw}. The four body decay distribution for $\bar{B_s} \to K^{*+}(\to K\pi) \ell^- \bar{\nu}_\ell$ can be fully characterized by four kinematical variables : the invariant mass squared of the dilepton $(q^2)$ and three angle variables $\theta_\ell, \theta_{K^*}$ and $\phi$. 
\begin{figure}[htb!]
\includegraphics[height = 5cm, width=0.7\textwidth]{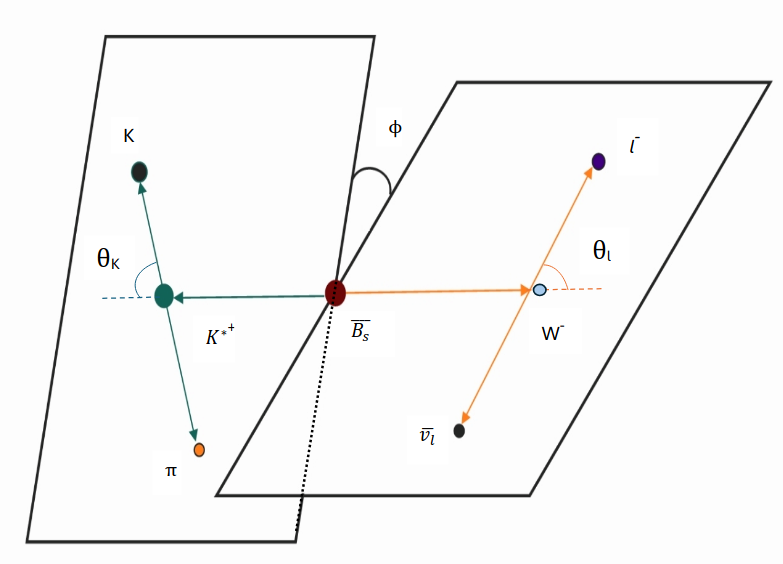}
\caption{The kinematics of the semileptonic decay $\bar{B_s} \to K^{*+}(\to K\pi) \ell^- \bar{\nu}_\ell$}
\end{figure}
Here, $\theta_\ell$ is the angle between $\ell$ in dilepton rest frame and the direction opposite to the B meson, $\theta_K$ is the angle between $K$ meson in the $K^*$ rest frame and the direction opposite to the B meson and $\phi$ is the angle between the plane defined by the dilepton and the plane defined by the $K\pi$ pair.

The fourfold differential distribution for this decay is given by

\begin{eqnarray}
\frac{d^4\Gamma}{dq^2\,d\cos\theta_\ell\, d\cos\theta_{K^*}d\phi} &\,=\, \frac{3}{8\pi}\,|N|^2\,\Big[ (J_{1s}+J_{2s}\ctl+ J_3 \sstl\cos\,2\phi + J_{6s}\cos\theta_\ell \nonumber\\
& + J_9 \sstl\sin\, 2\phi )\sstk +( J_{1c} +  J_{2c}\ctl)\cstk \nonumber \\
&+( J_4 \stl\cos\phi + J_5\sin{\theta_\ell}\cos\phi + J_7\sin\theta_\ell\sin\phi \nonumber\\
& + J_8 \stl \sin\phi )\stk +  J_{6c}\cstk\cos\theta_\ell\Big] 
\label{diff_eqn}
\end{eqnarray}
The coefficients $J_i(q^2)$ are the function of the dilepton invariant mass square and encapsulate the dynamic of the decay. These coefficients contain the form factors and are sensitive to different new physics. We have assumed the massless lepton limit. This approximation is valid for the electron channel and remains reliable for the muon channel, except in low $q^2$ regions where $m_\mu^2/q^2$ effects become non-negligible. In the massless limit of lepton, these coefficients can be written as:
\begin{eqnarray}
J_{1s} &=& \frac{3}{16}\Big[3 {|\mathcal{A}^L_\perp|}^2 + 3 {|\mathcal{A}^L_\parallel|}^2 + 16 {|\mathcal{A}_{0\parallel}|}^2 + 16 {|\mathcal{A}_{t\perp}|}^2  \Big] \nonumber\\
J_{1c} &=& \frac{3}{4}\Big[{|\mathcal{A}^L_{0}|}^2 +2 {|\mathcal{A}_{t}^L|}^2  +8 {|\mathcal{A}_{\parallel\perp}|}^2  \Big] \nonumber\\
J_{2s} &=& \frac{3}{16}\Big[{|\mathcal{A}^L_\perp|}^2+{|\mathcal{A}^L_\parallel|}^2 -16 {|\mathcal{A}_{0\parallel}|}^2 -16 {|\mathcal{A}_{t\perp}|}^2  \Big] \nonumber\\
J_{2c} &=& -\frac{3}{4}\Big[{|\mathcal{A}^L_{0}|}^2 -8 {|\mathcal{A}_{\parallel\perp}|}^2  \Big] \nonumber\\
J_{3} &=& \frac{3}{8}\Big[{|\mathcal{A}^L_\perp|}^2- {|\mathcal{A}^L_\parallel|}^2+ 16 {|\mathcal{A}_{0\parallel|}}^2 - 16 {|\mathcal{A}_{t\perp}|}^2  \Big] \nonumber\\
J_{4} &=& \frac{3}{4\sqrt{2}}\Big[\mathcal{A}^L_{0}{\mathcal{A}^L_\parallel}^* - 8\sqrt{2}\mathcal{A}_{\parallel\perp}{\mathcal{A}_{0\parallel}}^* \Big] \nonumber\\
J_{5} &=& \frac{3}{2\sqrt{2}}Re\Big[\mathcal{A}^L_{0}\mathcal{A}^L_\perp +2\sqrt{2} \mathcal{A}_{0\parallel}{\mathcal{A}^L_t}^* \Big] \nonumber\\
J_{6s} &=& \frac{3}{2}Re\Big[\mathcal{A}^L_\parallel{\mathcal{A}^L_\perp}^* \Big] , \quad \quad
J_{6c} = -6 Re\Big[\mathcal{A}_{\parallel\perp}|{\mathcal{A}^L_t}^* \Big] \nonumber\\
J_{7} &=& \frac{3}{2\sqrt{2}}Im\Big[\mathcal{A}^L_{0} {\mathcal{A}^L_\parallel}^* - 2\sqrt{2}\mathcal{A}_{t\perp}{\mathcal{A}^L_t}^*\Big] \nonumber\\
J_{8} &=& \frac{3}{4\sqrt{2}}Im\Big[\mathcal{A}^L_{0}{\mathcal{A}^L_\perp}^*  \Big], \quad\quad
J_{9} = \frac{3}{4}Im\Big[\mathcal{A}^L_\perp{\mathcal{A}^L_\parallel}^* \Big]
\end{eqnarray}

The form factors of $B_s \to K^*$ are defined in appendix \ref{form_factors}, and helicity amplitudes are given in appendix \ref{hel_amp}.
The differential branching fraction $d\mathcal{B}/dq^2$ can be derived from equation \ref{diff_eqn} by integrating over the three angles in the ranges $\theta_\ell \in [0,\pi], \theta_{K^*} \in [0, \pi]$ and $\phi \in [0, 2\pi]$
\begin{equation}
\frac{d\mathcal{B}}{dq^2} = \tau_{B_s} |N|^2\Big[2J_{1s}+J_{1c} - \frac{1}{3}\Big(2J_{2s} + J_{2c}\Big)\Big]
\end{equation}
where the normalization factor is defined as
\begin{equation}
|N|^2 = \frac{1}{3}\frac{G_F^2\, V_{ub}^2\, q^2 \sqrt{\lambda}}{2^{10}\pi^3 M_{B_s}^3}
\label{norm}
\end{equation}
with $\lambda = \lambda(M_{B_s}^2, M_{K^*}^2, q^2)$ which is the usual kinematic Kallen function. 
The forward-backward asymmetry for lepton can be defined as
\begin{equation}
A_{FB} = \frac{\int_0^1 \frac{d^2\Gamma}{dq^2\, d\cos\theta_\ell}d\cos\theta_\ell - \int_{-1}^0 \frac{d^2\Gamma}{dq^2\, d\cos\theta_\ell}d\cos\theta_\ell}{\int_{-1}^1 \frac{d^2\Gamma}{dq^2\, d\cos\theta_\ell}d\cos\theta_\ell}
\end{equation}

After simplifying the above expression, this asymmetry can be expressed as given below
\begin{equation}
A_{FB} =  \frac{J_{6s} + \frac{1}{2}J_{6c}}{\Big[2J_{1s}+J_{1c} - \frac{1}{3}\Big(2J_{2s} + J_{2c}\Big)\Big]}
\end{equation}

The longitudinal polarization of $K^*$ meson can be written as :
\begin{equation}
F_L  = \frac{3\Big(J_{1c} - \frac{1}{3}J_{2c}\Big)}{3(2 J_{1s} + J_{1c}) - (2 J_{2s} + J_{2c})}
\end{equation} 
To avoid dependence on overall normalization (CKM factors, hadronic uncertainties), we define the normalized angular observables as
\begin{equation}
\tilde{J}_i(q^2) = \frac{ |N|^2 \, J_i(q^2)}{\frac{d\Gamma}{dq^2}}
\end{equation}
where $|N|^2$ is the normalization factor defined in equation \ref{norm}. By construction, the normalized observables $\tilde{J}_i$ are independent of the CKM matrix element $V_{ub}$ and provide clean probes of new physics.

\section{Predictions of observables in $\bar{B_s} \to K^{*+}(\to K\pi) \ell^- \bar{\nu}_\ell$}
\label{predictions}
We explore the effects of NP WCs on the observables in the semileptonic decay $\bar{B_s} \to K^{*+}(\to K\pi) \ell^- \bar{\nu}_\ell$. We predict the observables differential branching ratio $\frac{d\mathcal{B}}{dq^2}$, lepton forward-backward asymmetry $A_{FB}$, $K^{*}$ longitudinal polarization $F_{L}$ and  normalized angular observables $\tilde{J}_i$'s using the best fit points of tables \ref{best-fit_1d} and \ref{best-fit_2d}. For the observables $\frac{d\mathcal{B}}{dq^2}, A_{FB} \, \text{and}\, F_L$, we present predictions only for those NP scenarios that show significant deviations from the SM expectations. In contrast, predictions for the angular observables $J_i$'s are provided for all the 1D and 2D NP scenarios considered in our analysis.  The error band in SM is computed by using the $1\sigma$ uncertainties in the hadronic form factors, together with CKM element error. We have considered the best fit points for the NP scenarios, and the $1\sigma$ uncertainties in the hadronic form factors, together with CKM element error, are included for NP predictions. To estimate the theoretical uncertainties on the observables, we adopt a one-sigma variation approach. Each input parameter, including the hadronic form factor parameters and the CKM element $V_{ub}$, is varied independently by $\pm 1$ standard deviation from its central value. The resulting variations in the observables are used to determine their one-sigma theoretical uncertainties.
We can summarize the potential evidence of new physics in the predictions of the different observables:
\begin{itemize}
\item {\bf Differential branching fraction $\left(\frac{d\mathcal{B}}{dq^2}\right)$ :} The differential branching fraction prediction plots for the three different scenarios $S6, S7 \& \,S17 \,$ are given in Figure. \ref{pred_plot_dbr}. The SM band is shown in green color where the band is due to the errors in CKM and form factor parameters. 
We find that all these scenarios can be distinguished from the SM as these scenarios provide the deviations from the SM prediction. We cannot distinguish these NP scenarios from each other, as they provide almost same type of deviation from SM. Also, the four benchmark points for scenario $S7$ provides the almost same $q^2$ spectrum as shown in the middle panel of FIG. \ref{pred_plot_dbr}.
\begin{figure}[htb!]
\includegraphics[height=0.22\textwidth,width=0.3\textwidth]{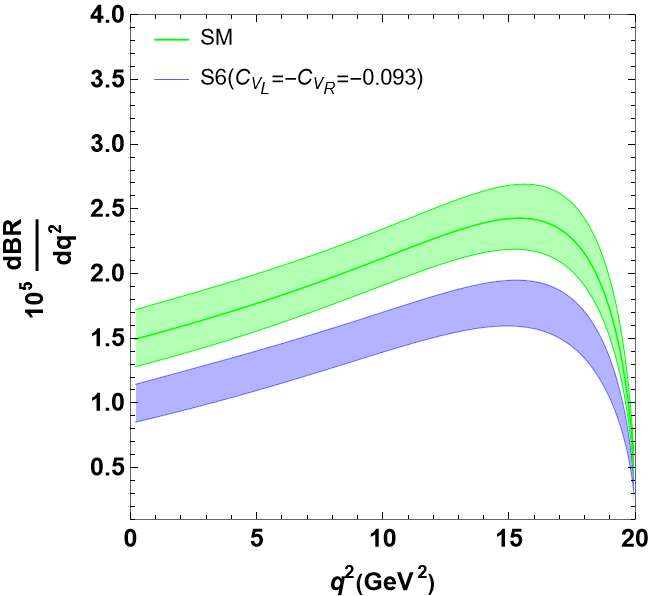}
\includegraphics[height=0.22\textwidth,width=0.3\textwidth]{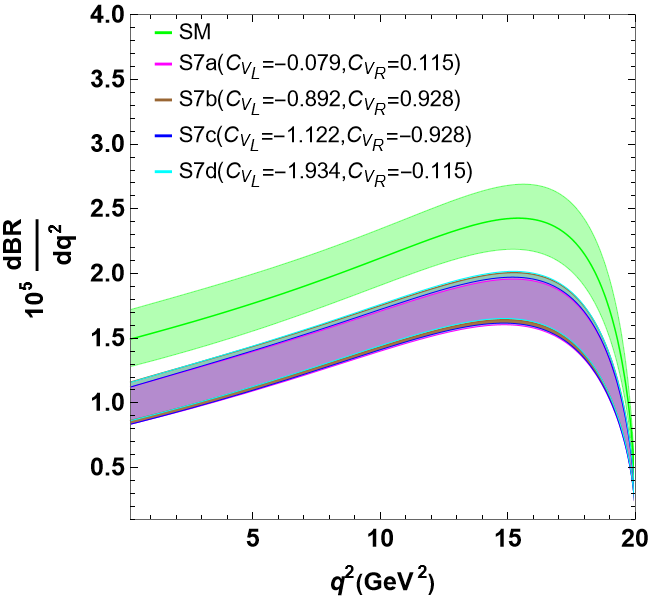}
\includegraphics[height=0.22\textwidth,width=0.3\textwidth]{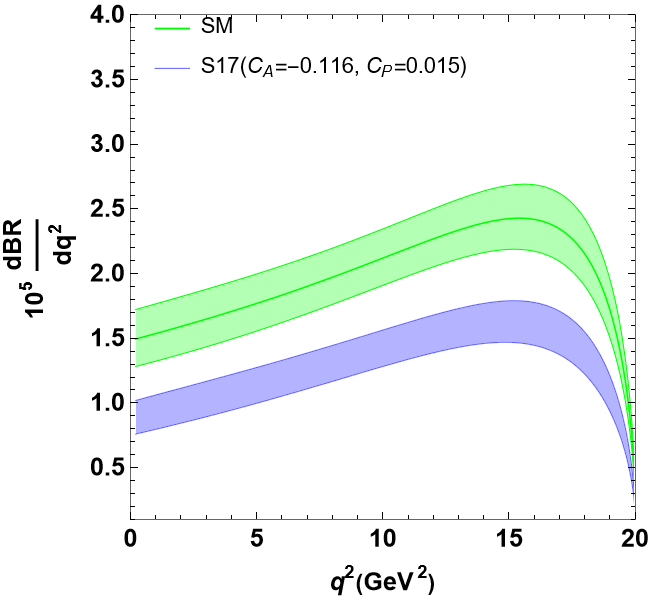}
\caption{The $q^2$ dependence of $dB/dq^2$ in $\bar{B_s} \to K^{*+}(\to K\pi) \ell^- \bar{\nu}_\ell$. The SM band is given in green color while other bands with different colors correspond to different NP benchmark points. The error band is computed by using the $1\sigma$ uncertainties in the hadronic form factors, together with CKM element error.}
\label{pred_plot_dbr} 
\end{figure}

\item {\bf Leptonic forward-backward asymmetry $(A_{FB})$ :} The predictions for lepton forward-backward asymmetry are shown in Figure. \ref{pred_plot_afb}. The NP scenarios $S6 : C_{V_L} = -C_{V_R}$ and $S17 : (C_{V_L} = -C_{V_R}, C_{S_L} = -C_{S_R})$ provide small deviations from the SM in the forward-backward asymmetry as shown in the left and right panels of FIG. \ref{pred_plot_afb}. 
\begin{figure}[htb!]
\includegraphics[height=0.22\textwidth,width=0.3\textwidth]{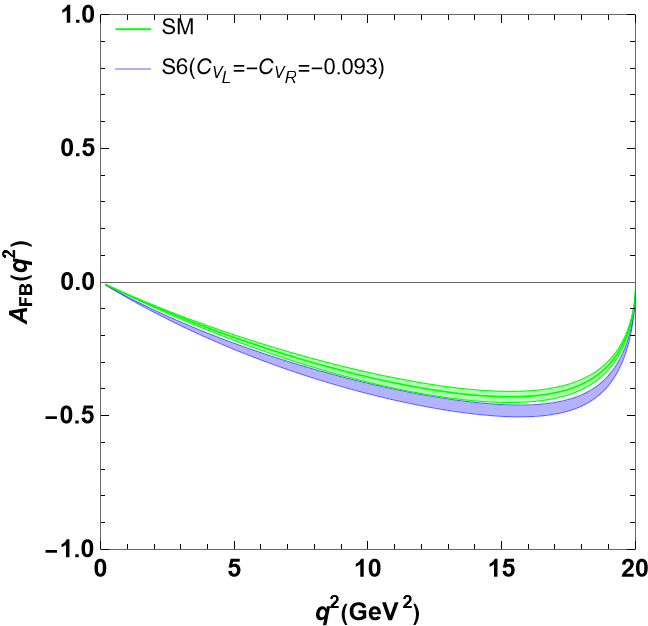}
\includegraphics[height=0.22\textwidth,width=0.3\textwidth]{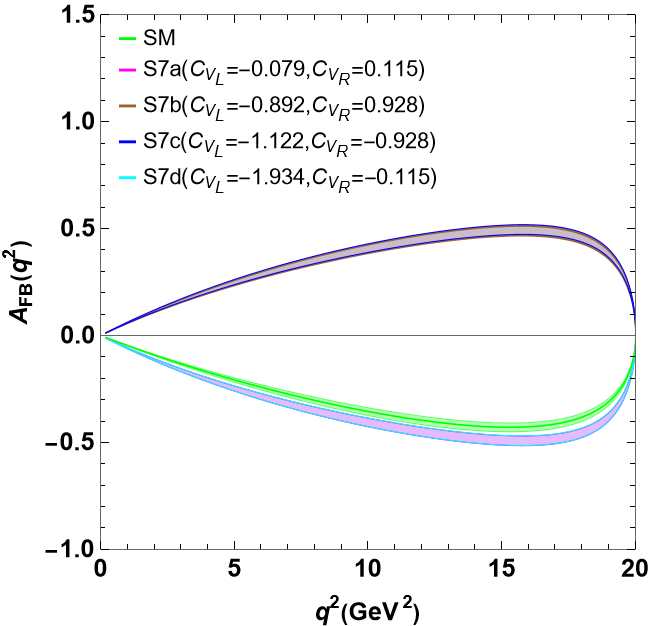}
\includegraphics[height=0.22\textwidth,width=0.3\textwidth]{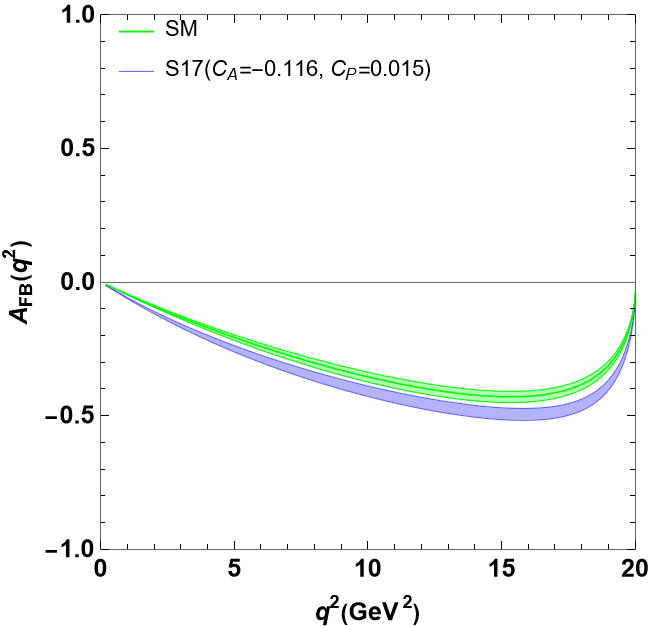}
\caption{The $q^2$ dependence of Lepton forward-backward asymmetry $A_{FB}$ in $\bar{B_s} \to K^{*+}(\to K\pi) \ell^- \bar{\nu}_\ell$. The SM band is given in green color while other bands with different colors correspond to different NP benchmark points. The error band is computed by using the $1\sigma$ uncertainties in the hadronic form factors, together with CKM element error.}
\label{pred_plot_afb} 
\end{figure}

The two benchmark points (BPs) $S7a\,\&\, S7d$ for scenario $S7 : (C_{V_L}, C_{V_R})$ provide a small deviation from the SM, but the other two scenarios $S7b\,\&\, S7c$ provide a large deviation in the forward-backward asymmetry (shown in the middle panel of FIG. \ref{pred_plot_afb}). The BPs $S7a\,\&\, S7d$ provide the negative $A_{FB}$ for the whole $q^2$ range, while $S7b\,\&\, S7c$ BPs give the positive $A_{FB}$ for the full $q^2$ range. The benchmark points $S7b\,\&\, S7c$ can be distinguished from the other two benchmark points $S7a\, \&\, S7d$ based on the forward-backward asymmetry which were indistinguishable from the differential branching fraction $\frac{d\mathcal{B}}{dq^2}$.

\item {\bf Longitudinal polarization of $K^*$ meson $(F_L)$ :} The prediction plots for longitudinal polarization fraction of $K^*$ meson are shown in Figure. \ref{pred_plot_fl}.
\begin{figure}[htb!]
\includegraphics[height=0.22\textwidth,width=0.3\textwidth]{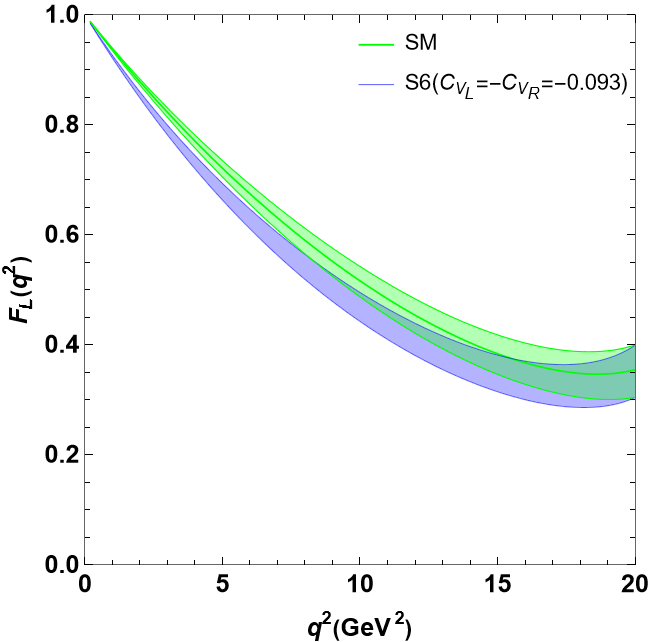}
\includegraphics[height=0.22\textwidth,width=0.3\textwidth]{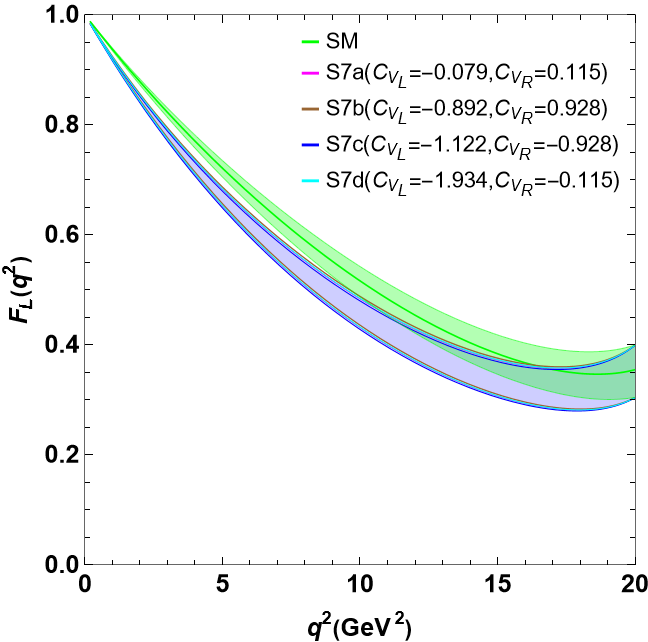}
\includegraphics[height=0.22\textwidth,width=0.3\textwidth]{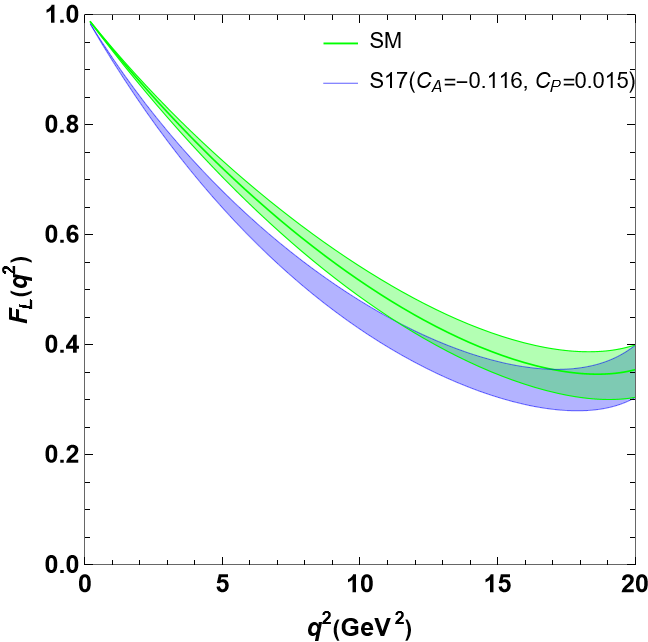}
\caption{The $q^2$ dependence of the longitudinal polarization of $K^*$ meson $(F_L)$ in $\bar{B_s} \to K^{*+}(\to K\pi) \ell^- \bar{\nu}_\ell$ decay. The SM band is given in green color while other bands with different colors correspond to different NP benchmark points.  The error band is computed by using the $1\sigma$ uncertainties in the hadronic form factors, together with CKM element error.}
\label{pred_plot_fl} 
\end{figure}
The NP scenarios provide a small deviation in the longitudinal polarization of $K^*$ meson. We find almost similar predictions for the longitudinal polarization asymmetry for all three scenarios. It is not possible to differentiate the NP scenarios with this observable.

\item {\bf Predictions of normalized angular observables $(\tilde{J_i})$ :} The $q^2$ spectra for the normalized angular observables are shown in Figure~\ref{pred_plot_J1d} for 1D NP scenarios. We observe that the normalized angular observables $\tilde{J}_{1c}$, $\tilde{J}_{1s}$, $\tilde{J}_{2c}$, and $\tilde{J}_{2s}$ exhibit very limited sensitivity to the 1D NP scenarios. In contrast, the observables $\tilde{J}_3$, $\tilde{J}_4$, $\tilde{J}_5$, and $\tilde{J}_{6s}$ show significant sensitivity to the two NP scenarios, $S2 (C_{V_R})$ and $S6 (C_{V_L} = -C_{V_R})$, with all four exhibiting similar deviations from the SM predictions.
\begin{figure}[htb!]
\includegraphics[height=0.22\textwidth,width=0.3\textwidth]{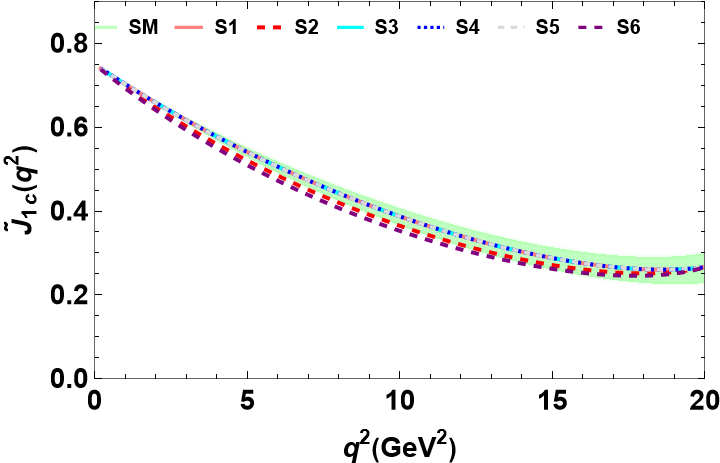}
\includegraphics[height=0.22\textwidth,width=0.3\textwidth]{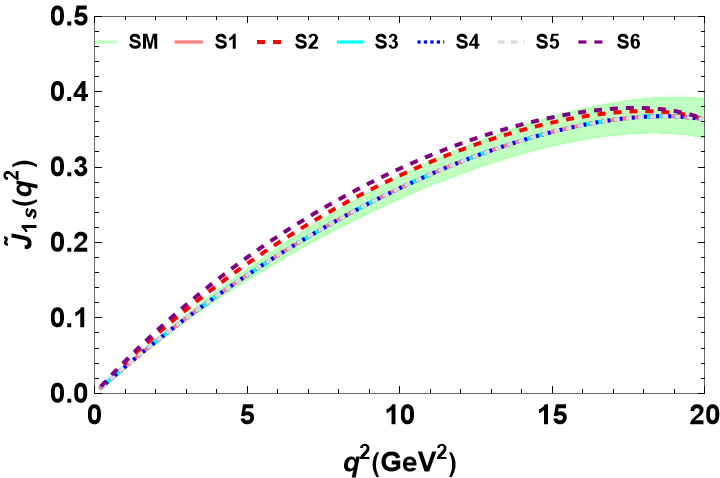}
\includegraphics[height=0.22\textwidth,width=0.3\textwidth]{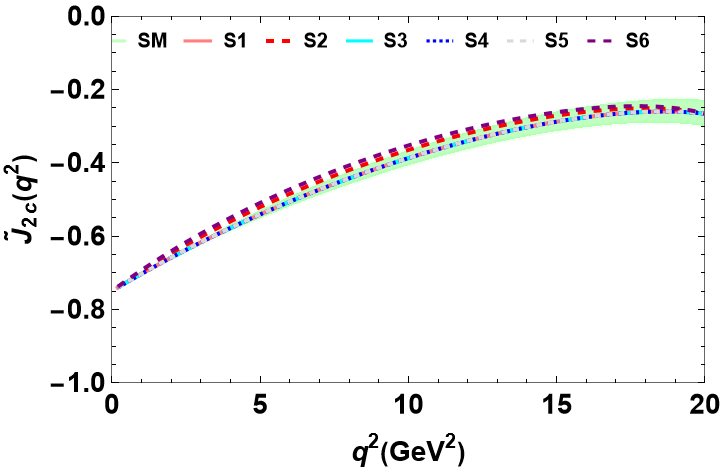}\\
\includegraphics[height=0.22\textwidth,width=0.3\textwidth]{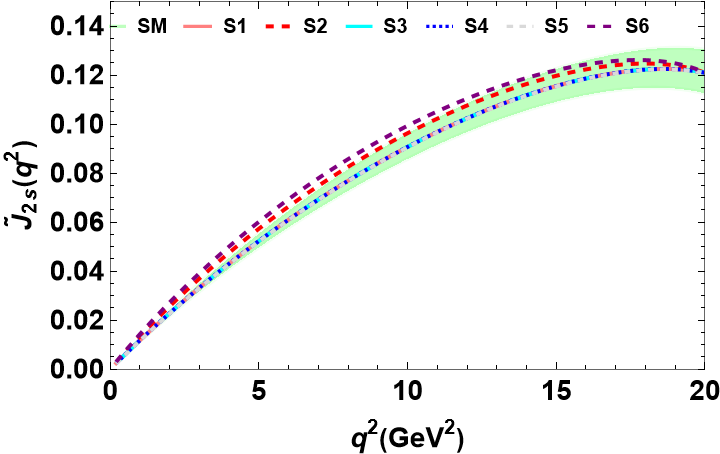}
\includegraphics[height=0.22\textwidth,width=0.3\textwidth]{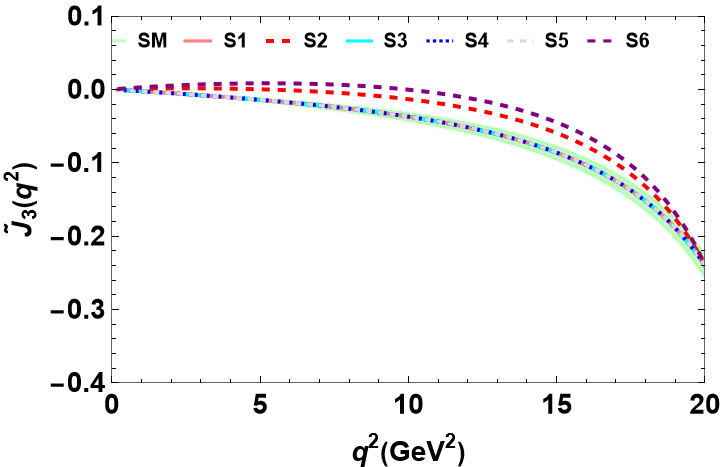}
\includegraphics[height=0.22\textwidth,width=0.3\textwidth]{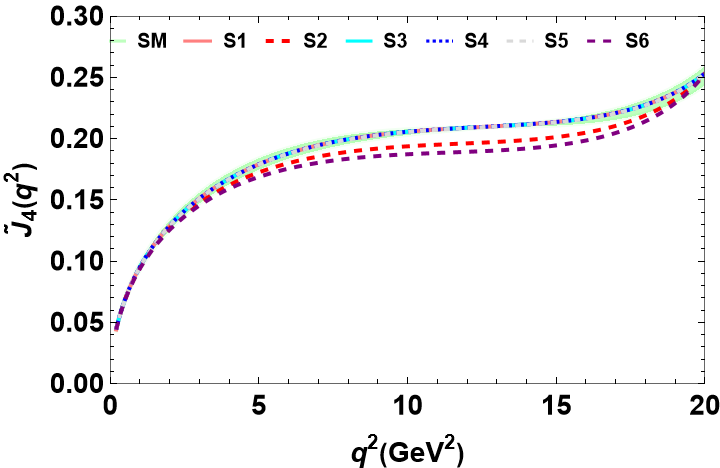}\\
\includegraphics[height=0.22\textwidth,width=0.3\textwidth]{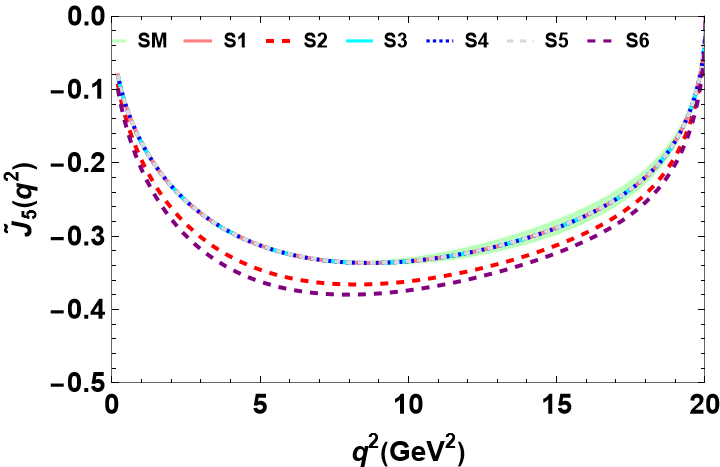}
\includegraphics[height=0.22\textwidth,width=0.3\textwidth]{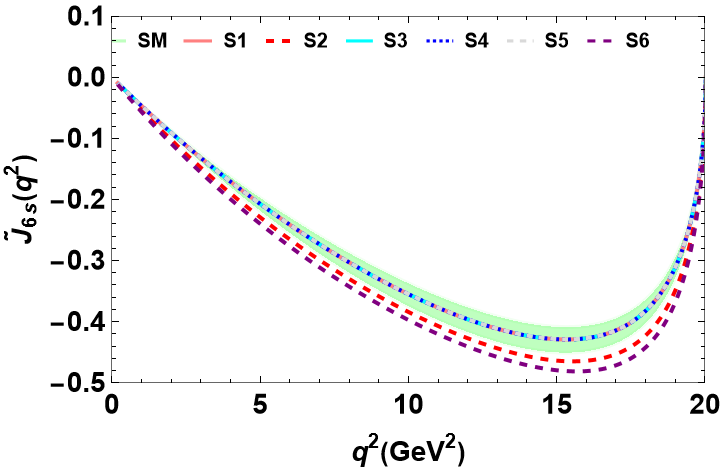}
\caption{The $q^2$ dependence of the angular observables in $\bar{B_s} \to K^{*+}(\to K\pi) \ell^- \bar{\nu}_\ell$ decay in 1D NP scenarios. The error band is computed by using the $1\sigma$ uncertainties in the hadronic form factors, together with CKM element error. The benchmark points for different scenarios are given as : $S1 : (C_{V_L}  = -0.032), S2 : (C_{V_R} = 0.069), S3 : (C_{S_L} = -0.003), S4 : (C_{S_R} = 0.003), S5 : (C_T = 0.005), S6 : (C_{V_L} = -C_{V_R} = -0.093)$}
\label{pred_plot_J1d} 
\end{figure}

The $q^2$ dependence of normalized angular observables for 2D NP scenarios are shown in Figure~\ref{pred_plot_J2d}. The normalized angular observables $\tilde{J}_{1c}$, $\tilde{J}_{1s}$, $\tilde{J}_{2c}$, and $\tilde{J}_{2s}$ exhibit the NP sensitivity for 2D scenarios S7$(C_{V_L}, C_{V_R})$ and S17$(C_{V_L} = -C_{V_R}, C_{S_L} = -C_{S_R})$. The observables $\tilde{J}_3$, $\tilde{J}_4$, $\tilde{J}_5$, and $\tilde{J}_{6s}$ show a significant deviation from SM for 2D NP scenarios S11$(C_{V_R}, C_{S_L})$, S12$(C_{V_R}, C_{S_R})$ $\&$ S13$(C_{V_R}, C_{T})$, while these observables are very sensitive for the 2D NP scenarios S7$(C_{V_L}, C_{V_R})$ and S17$(C_{V_L} = -C_{V_R}, C_{S_L} = -C_{S_R})$. These observables can be very helpful to check the NP sensitivity, but it is very challenging to measure these  observables in near future at LHCb.

\begin{figure}[htb!]
\includegraphics[height=0.22\textwidth,width=0.3\textwidth]{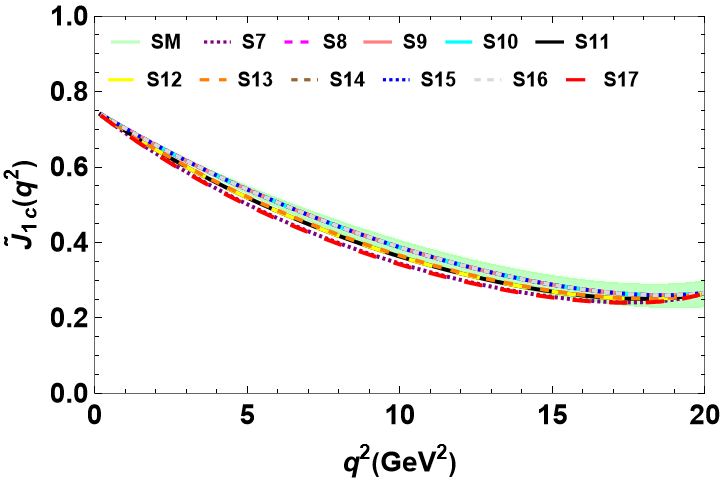}
\includegraphics[height=0.22\textwidth,width=0.3\textwidth]{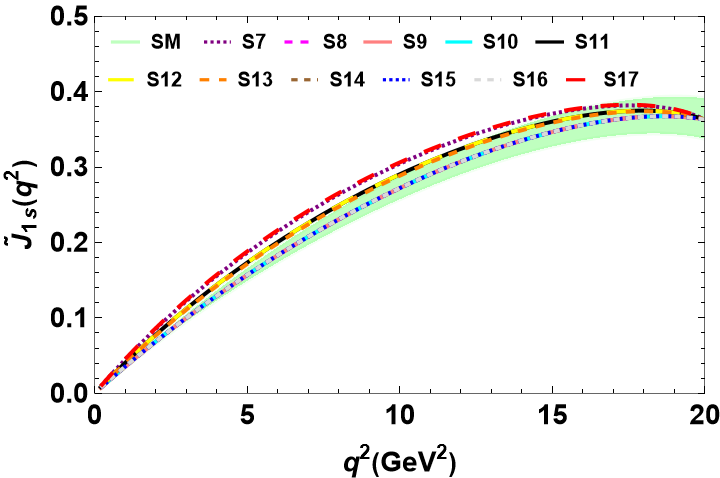}
\includegraphics[height=0.22\textwidth,width=0.3\textwidth]{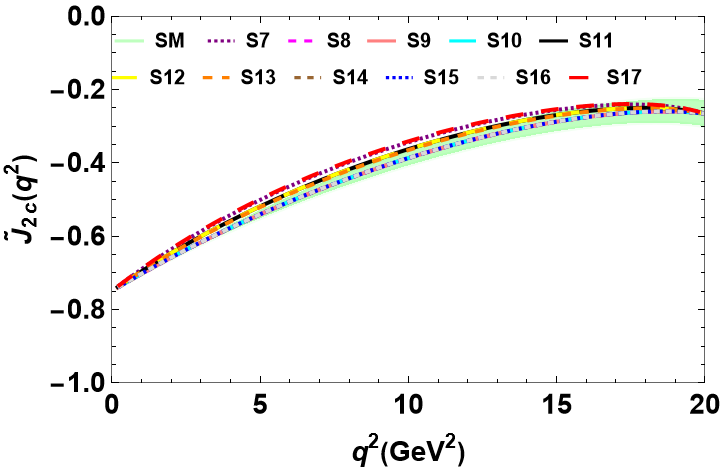} \\
\includegraphics[height=0.22\textwidth,width=0.3\textwidth]{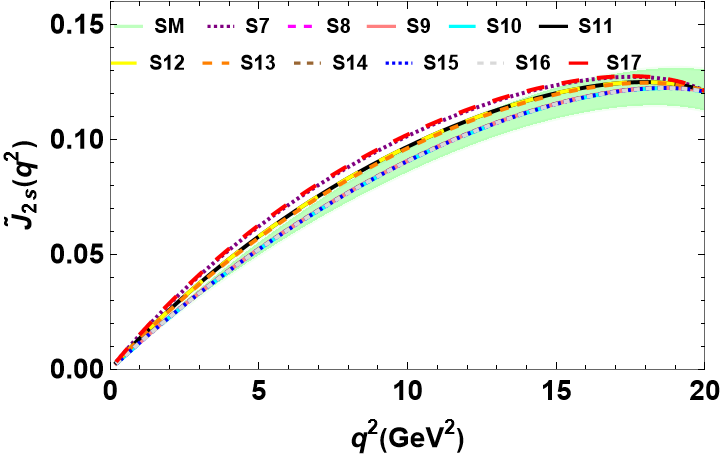}
\includegraphics[height=0.22\textwidth,width=0.3\textwidth]{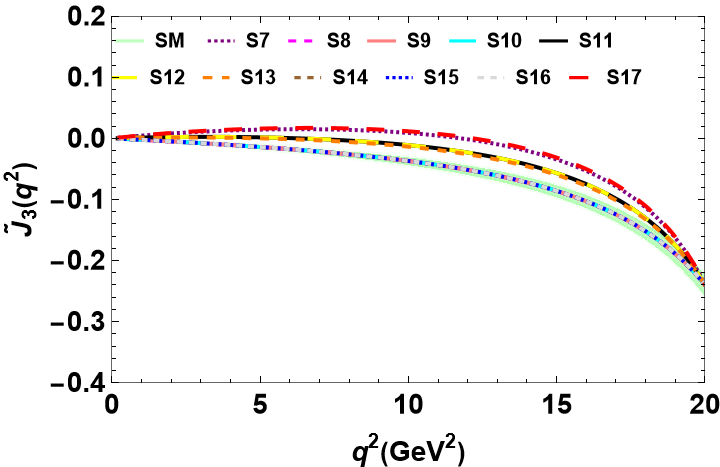}
\includegraphics[height=0.22\textwidth,width=0.3\textwidth]{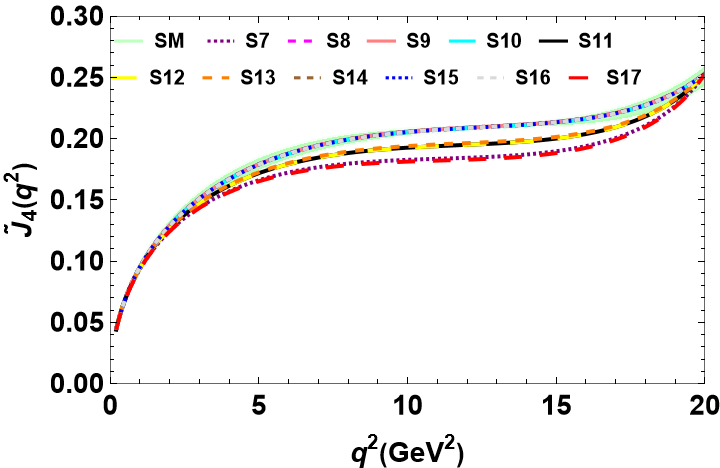} \\
\includegraphics[height=0.22\textwidth,width=0.3\textwidth]{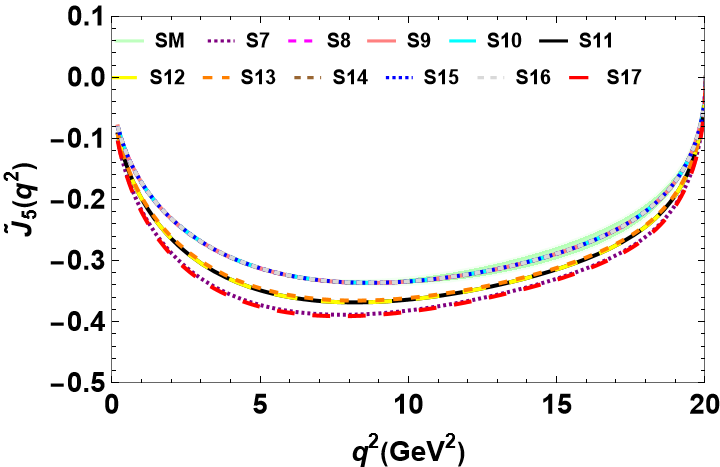}
\includegraphics[height=0.22\textwidth,width=0.3\textwidth]{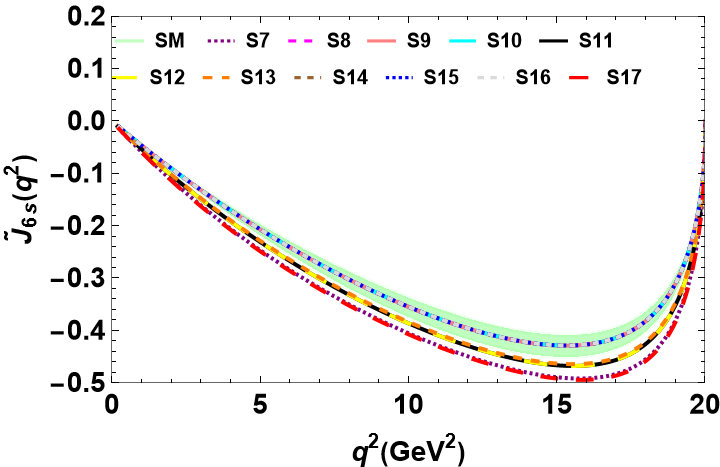}
\caption{The $q^2$ dependence of the angular observables in $\bar{B_s} \to K^{*+}(\to K\pi) \ell^- \bar{\nu}_\ell$ decay in 2D NP scenarios. The error band is computed by using the $1\sigma$ uncertainties in the hadronic form factors, together with CKM element error.  The benchmark points for different scenarios are given as : $S7 : (C_{V_L} = -0.079,C_{V_R} = 0.115),\,\, S8 : (C_{V_L} = -0.038,C_{S_L} = -0.003),\,\, S9 : (C_{V_L} = -0.038,C_{S_R} = 0.004),\,\, S10  : (C_{V_L} = -0.032,C_{T} = 0.006),\,\, S11 : (C_{V_R} = 0.075,C_{S_L}=-0.004),\,\, S12 : (C_{V_R}=0.075,C_{S_R}=0.004),\,\, S13 : (C_{V_R}=0.068,C_{T}=0.0007),\,\, S14 : (C_{S_L}=0.008,C_{S_R}=0.011),\,\, S15 : (C_{S_L}=-0.003,C_{T}=0.005),\,\, S16 : (C_{S_R}=0.003,C_{T}=0.005),\, \, S17 : (C_A = -0.116, C_P = 0.015)$}
\label{pred_plot_J2d} 
\end{figure}
\end{itemize}

\section{Conclusions}
\label{summary}
The Standard Model provides the elegant description of fundamental particles and their interaction. Any new particle beyond the SM is not observed yet, which can provide the direct evidence of NP. However, the measurements in $b \to c \ell \nu$ provided a hint about NP beyond the SM. The NP interactions in $b \to c \ell \nu$ can explain these discrepancies. We have analyzed the new physics description in $b \to u \ell \nu$ by considering the new physics in light leptons. The data in $b \to u \ell \nu$ sector is expected from the LHCb run-3 and Belle II experiments, so it is very useful to study the NP effects in $b \to u \ell \nu$.

In this work, we consider the model independent approach to inspect the NP in leptonic and semileptonic decays of $B$ mesons induced by the $b \to u \ell \nu_{\ell}$ transition. We work in effective field theory by considering the general effective Hamiltonian with different NP Lorentz structures. We consider the NP operators with one at a time and two at a time scenarios. The different NP Wilson coefficients in this analysis are constrained by using the available measurements in the $b \to u \ell\nu_{\ell}$ sector. The best fit of NP WCs are obtained by minimizing the $\chi^2$ function for both 1D and 2D scenarios. We also obtain the $1 \sigma$ and $2 \sigma$ contours for three 2D scenarios $S7 : (C_{V_L}, C_{V_R}), \,S14 : (C_{S_L}, C_{S_R})$ and $ S17 : (C_{V_L} = -C_{V_R}, C_{S_L} = -C_{S_R})$.

The NP WCs are used to predict the possible departure from SM of observables in $\bar{B_s} \to K^{*+}(\to K\pi) \ell^- \bar{\nu}_\ell$. We give the predictions of the $q^2$ spectrum of differential branching ratio, leptonic forward-backward asymmetry, longitudinal polarization of the $K^*$ meson and normalized angular observables in the semileptonic decay $\bar{B_s} \to K^{*+}(\to K\pi) \ell^- \bar{\nu}_\ell$ decay. Any deviations in the observables from the SM can indicate the presence of NP Lorentz structure. The NP scenarios $S6, S7$ and $S17$ decrease the differential $q^2$ spectrum from the SM. All these scenarios provide almost similar deviation in the branching ratio so it is not possible to distinguish them individually, but can be distinguished from the SM. The forward-backward asymmetry of lepton with NP effects also deviates from the SM prediction and the two benchmark points $S7a\,\&\, S7d$ of $S7 : (C_{V_L}, C_{V_R})$ scenario can be distinguished from the other benchmark points $S7b\,\&\, S7c$ as these two give the positive value of $A_{FB}(q^2)$ in the full $q^2$ range whereas the other benchmark points provide the negative $A_{FB}$. The lepton polarization fraction of $K^*$ meson also deviates from the SM. The $q^2$ dependence of the angular observables is also provided for the 1D and 2D NP scenarios. Measurements of $\bar{B_s} \to K^{*+}(\to K\pi) \ell^- \bar{\nu}_\ell$ at upcoming runs of LHCb and Belle II could contribute to a better understanding of the flavor structure and help explore possible deviations from the Standard Model.
\section*{Acknowledgements}
The work of DK is supported by the SERB, Govt. of India under the research grant no. SERB/EEQ/2021/000965. DK would like to thank Shireen Gangal and Jacky Kumar for useful discussion during this work. We honor the memory of Prof. Ashutosh Kumar Alok, whose guidance and discussion were essential to this project.

\begin{appendices}
  \renewcommand{\thesection}{} 
  \section{Appendix}
  \renewcommand{\thesubsection}{\Alph{subsection}}
\subsection{$B_s \to K^*$ transition form factors}
\label{form_factors}
The hadronic matrix elements for $B_s \to K^*$ can be written in terms of seven form factors, namely $V, A_0, A_1, A_{12}, T_1,T_2 $ and $T_{23}$. The form factors are defined by simplified series expansion in $z$ given by Bharucha-Straub-Zwicky \cite{Bharucha:2015bzk} as
\begin{equation}
f_i(q^2) = \frac{1}{(1-q^2/m_{R,i}^2)}\sum_k \alpha_k^i[z(q^2) - z(0)]^k
\end{equation}
where $z(q^2)$ is the conformal mapping variable and is defined as
\begin{equation}
z(t) = \frac{\sqrt{t_+ - t} - \sqrt{t_+ - t_0}}{\sqrt{t_+ - t} + \sqrt{t_+ - t_0}}
\end{equation}
with $t_{\pm} = (m_{B_s} \pm m_{K^*})$ and $t_0 = (m_{B_s}+m_{K^*})(\sqrt{m_{B_s}} - \sqrt{m_{K^*}})^2$. 
The fit results for the SSE expansion coefficients using the combined LCSR + Lattice fit and masses of sub-threshold resonances are provided in Ref. \cite{Bharucha:2015bzk}. We have summarized these resonance masses and SSE expansion coefficients in table \ref{res_masses} and table \ref{sse_coeff}, respectively.
\begin{table}[htb!]
\centering
\begin{tabular}{ |c|c|c|}
\hline
$f_i$ & $J^P$ & $m_{R,i}/GeV$ \\
\hline
$A_0$ & $0^-$& 5.279\\
\hline
$V, T_1$ & $1^-$ & 5.325\\
\hline
$A_1, T_2, A_{12}, T_{23}$ &$1^+$ & 5.724\\
\hline
\end{tabular}
\caption{Masses of resonances required for form factor parameterizations \cite{Bharucha:2015bzk}}
\label{res_masses}
\end{table}

\begin{table}[h!]
\centering
\begin{tabular}{ |c|c|c|c| }\hline
$f_i$ & $\alpha_0^i$ & $\alpha_1^i$ & $\alpha_2^i$ \\\hline
$V$ & $0.28 \pm 0.02$  & $-0.82 \pm 0.19$ & $5.08 \pm 1.42$ \\\hline
$A_0$ & $0.36 \pm 0.02$  & $-0.36 \pm 0.20$  & $8.03 \pm 2.07$ \\\hline
$A_1$ & $0.22 \pm 0.01$ & $0.24 \pm 0.16$  & $1.77 \pm 0.85$ \\\hline
$A_{12}$ & $0.27 \pm 0.02$  & $1.12 \pm 0.11$   & $3.43 \pm 0.78$ \\\hline
$T_1$  &$0.24 \pm 0.01$ & $-0.75 \pm 0.15$ & $2.49 \pm 1.37$ \\\hline
$T_2$ & $0.24 \pm 0.01$& $0.31 \pm 0.15$& $1.58 \pm 0.93$ \\\hline
$T_{23}$ & $0.60 \pm 0.04$ & $2.40 \pm 0.27$ & $9.64 \pm 2.03$\\\hline
\end{tabular}
\caption{Simplified series expansion coefficients $\alpha_k^i$ for parameterizing the $B_s \to K^*$ form factors \cite{Bharucha:2015bzk}}
\label{sse_coeff}
\end{table}

The form factors in the helicity basis can be written as \cite{Feldmann:2015xsa}
\setlength{\abovedisplayskip}{0pt} \setlength{\belowdisplayskip}{0pt} \setlength{\abovedisplayshortskip}{0pt} \setlength{\belowdisplayshortskip}{0pt}

\begin{itemize}
\item {Vector current}
\begin{equation}
\mathcal{F}_\perp(q^2) = \dfrac{\sqrt{2 \lambda}}{M_{B_s}(M_{B_s}+M_{K^*})} V(q^2)
\end{equation}

\item {Axial vector current}
\begin{eqnarray}
\mathcal{F}_t(q^2) &= \dfrac{\sqrt{\lambda}}{{M_{B_s}}^2} A_0(q^2) \\
\mathcal{F}_\parallel(q^2) &= \sqrt{2} \frac{M_{B_s}+M_{K^*}}{M_{B_s}} A_1(q^2)\\
\mathcal{F}_0(q^2) & = \dfrac{8M_{K^*} A_{12}(q^2) }{M_{B_s}}
\end{eqnarray}
\item{Tensor current}
\begin{eqnarray}
\mathcal{F}_{\perp}^T(q^2) &= \dfrac{\sqrt{2\lambda}}{{M_{B_s}}^2} T_1(q^2) \\
\mathcal{F}_{\parallel}^T(q^2) &= \dfrac{\sqrt{2}({M_{B_s}}^2-{M_{K^*}}^2)}{{M_{B_s}}^2} T_2(q^2) \\
\mathcal{F}^T_0(q^2) & = \dfrac{4 M_{K^*} T_{23}(q^2) }{M_{B_s}+M_{K^*}}
\end{eqnarray}
\end{itemize}

\subsection{Helicity amplitudes}
\label{hel_amp}
The helicity amplitudes for $\bar{B_s} \to K^{*+}(\to K\pi) \ell^- \bar{\nu}_\ell$ are given as \cite{Feldmann:2015xsa}
\begin{align}
\mathcal{A}_0^L &= -4\dfrac{M_{B_s}^2(1+C_{V_L}-C_{V_R})\mathcal{F}_0(q^2)}{\sqrt{q^2}}\\
\mathcal{A}_{\perp}^L &= 4 M_{B_s}(1+C_{V_L}+C_{V_R})\mathcal{F}_\perp(q^2)\\
\mathcal{A}_{\parallel}^L &= -4 M_{B_s}(1+C_{V_L}-C_{V_R})\mathcal{F}_\parallel(q^2)\\
\mathcal{A}_t^L &= -4 \big[{ \dfrac{m_l M_{B_s}^2}{\sqrt{q^2}}}(1+C_{V_L}-C_{V_R}) +\dfrac{M_{B_s}^2}{m_b}(C_{S_L}-C_{S_R})\big]\mathcal{F}_t(q^2)\\
\mathcal{A}_{\parallel\perp} &= +8 M_{B_s} C_T \mathcal{F}_0^T(q^2)\\
\mathcal{A}_{t\perp} &= 4 \sqrt{2} \dfrac{M_{B_s}^2}{\sqrt{q^2}} C_T \mathcal{F}_{\perp}^{T}(q^2)\\
\mathcal{A}_{0\parallel} &= 4 \sqrt{2} \dfrac{M_{B_s}^2}{\sqrt{q^2}} C_T \mathcal{F}_{\parallel}^{T}(q^2)
\end{align}
\end{appendices}

\end{document}